\documentclass[10pt,journal,compsoc]{IEEEtran}

\ifCLASSOPTIONcompsoc

  \usepackage[nocompress]{cite}

\usepackage{amsmath}

\usepackage{algorithmic}

\usepackage{array}

\usepackage{dblfloatfix}

\usepackage{url}
\usepackage{booktabs}
\usepackage{tikz}
\usepackage{makecell}
\usepackage{threeparttable}
\usepackage{longtable}
\usepackage{multirow}
\usepackage{tabularx}
\usepackage{tabulary}
\usepackage{tablefootnote}
\usepackage{amssymb}
\usepackage{epstopdf}
\usepackage{graphicx}
\usepackage{subfig}
\usepackage{caption}
\usepackage{color}
\usepackage{mathtools}
\usepackage{hyperref}
\usepackage{bm}
\usepackage{dsfont}
\usepackage{enumitem}
\begin{document}

\title{A$^2$-GCN: An Attribute-aware Attentive GCN Model for Recommendation}

\author{Fan~Liu,
        Zhiyong~Cheng,
        Lei~Zhu,
        Chenghao~Liu,
	    and~Liqiang~Nie,~\IEEEmembership{Senior Member,~IEEE}
	\IEEEcompsocitemizethanks{\IEEEcompsocthanksitem Liqiang Nie and Zhiyong Cheng are the joint corresponding authors.}
	\IEEEcompsocitemizethanks{\IEEEcompsocthanksitem F. Liu and L. Nie are with School of Computer Science and Technology, Shandong University, China. Email: liufancs@gmail.com, nieliqiang@gmail.com}
	\IEEEcompsocitemizethanks{\IEEEcompsocthanksitem Z. Cheng is with the Shandong Artificial Intelligence Institute, Qilu University of Technology (Shandong Academy of Sciences), China.
	Email: jason.zy.cheng@gmail.com}
	\IEEEcompsocitemizethanks{\IEEEcompsocthanksitem L. Zhu is  with the School of Information Science and Engineering, Shandong Normal University, China. Email:leizhu0608@gmail.com}
	\IEEEcompsocitemizethanks{\IEEEcompsocthanksitem C. Liu is with  School of Computer Science and Technology, Zhejiang University, China, and with School of Information Systems, Singapore Management University, Singapore. Email:twinsken@zju.edu.cn}
    }

\markboth{IEEE TRANSACTIONS ON KNOWLEDGE AND DATA ENGINEERING}%
{Shell \MakeLowercase{\textit{et al.}}: Bare Demo of IEEEtran.cls for Computer Society Journals}

\IEEEtitleabstractindextext{
\begin{abstract}
As important side information, attributes have been widely exploited in the existing recommender system for better performance. In the real-world scenarios, it is common that some attributes of items/users are missing (e.g., some movies miss the genre data). Prior studies usually use a default value (i.e., ``other") to  represent the missing attribute, resulting in sub-optimal performance. To address this problem, in this paper, we present an attribute-aware attentive graph convolution network (A${^2}$-GCN). In particular, we first construct a graph, whereby users, items, and attributes are three types of nodes and their associations are edges. Thereafter, we leverage the graph convolution network to characterize the complicated interactions among $<$users, items, attributes$>$. To learn the node representation, we turn to the message-passing strategy to aggregate the message passed from the other directly linked types of nodes (e.g., a user or an attribute). To this end, we are capable of incorporating associate attributes to strengthen the user and item representations, and thus naturally solve the attribute missing problem. Considering the fact that for different users, the attributes of an item have different influence on their preference for this item, we design a novel attention mechanism to filter the message passed from an item to a target user by considering the attribute information. Extensive experiments have been conducted on several publicly accessible datasets to justify our model. Results show that our model outperforms several state-of-the-art methods and demonstrate the effectiveness of our attention method.
\end{abstract}

}

\maketitle

\IEEEdisplaynontitleabstractindextext

\IEEEpeerreviewmaketitle

\section{Introduction}
\IEEEPARstart{R}{ecommendation} has long been one of the core techniques for various platforms, such as E-commerce website, news portals and social media sites. It not only helps users find the content of interest from overwhelming information, but also increases the revenue for the service provider (e.g., Amazon\footnote{https://www.amazon.com.}, eBay\footnote{https://www.eBay.com.}, Mercari\footnote{https://www.mercari.com.\label{mecari}}). Collaborative filtering (CF) based methods~\cite{Koren2009MF,he2017neural}, like matrix factorization~\cite{Koren2009MF}, have achieved great success in learning user and item representations via modeling user-item interaction behavior. However, their performance is often unsatisfactory when the interactions are sparse. To tackle this issue, an effective solution is to leverage related side information (such as reviews, images, social relations, and attributes), which provides additionally rich information for user preference and item feature modeling~\cite{Rendle2011Factorization}. 

Among various side information, the most exploited one is the attribute. Typical examples include category and brand of products, genre and directors of movies, as well as descriptive tags of social images. Many methods have been developed to leverage above attributes in recommendation~\cite{Rendle2011Factorization,cheng2016wide,he2011NFM,shi2018accm}, because they provide important and concise information for items. In general, existing methods are mainly in two paradigms according to the ways of exploiting attributes. One is to convert attributes into binary feature vectors via one-hot/multi-hot encoding by generalized linear models~\cite{Rendle2011Factorization,cheng2016wide}, such as logistic regression, support vector machines, linear part of factorization machines (FMs)\cite{Rendle2011Factorization}, and Wide\&Deep~\cite{cheng2016wide}. Despite their success, their defect is that the features will become very sparse via one-hot/multi-hot encoding, making it difficult to learn reliable parameters~\cite{Rendle2011Factorization}. Besides, it also significantly increases the feature space. 
For instance, in Amazon’s Kindle Store dataset, there are 1,438 attribute labels in total. A 1,438-dimension feature vector will be generated via the multi-hot embedding method. In this feature vector, only few dimensions are non-zeros. Another way is to first embed each attribute into an $k$-dimension feature vector, and then concatenate the feature vectors of all the attributes as an input feature vector for subsequent modules~\cite{he2011NFM,shi2018accm,cheng2016wide}. For example, NFM~\cite{he2011NFM} projects each feature to a dense representation via the embedding layer. This approach is widely-adopted in recently proposed deep learning based models\cite{shi2018accm,cheng2016wide}, which can alleviate the aforementioned sparsity problem. However, there are still some shortcomings in the above two kinds of methods. 

\begin{figure}[ht]
	\centering
	\includegraphics[width=0.8\linewidth]{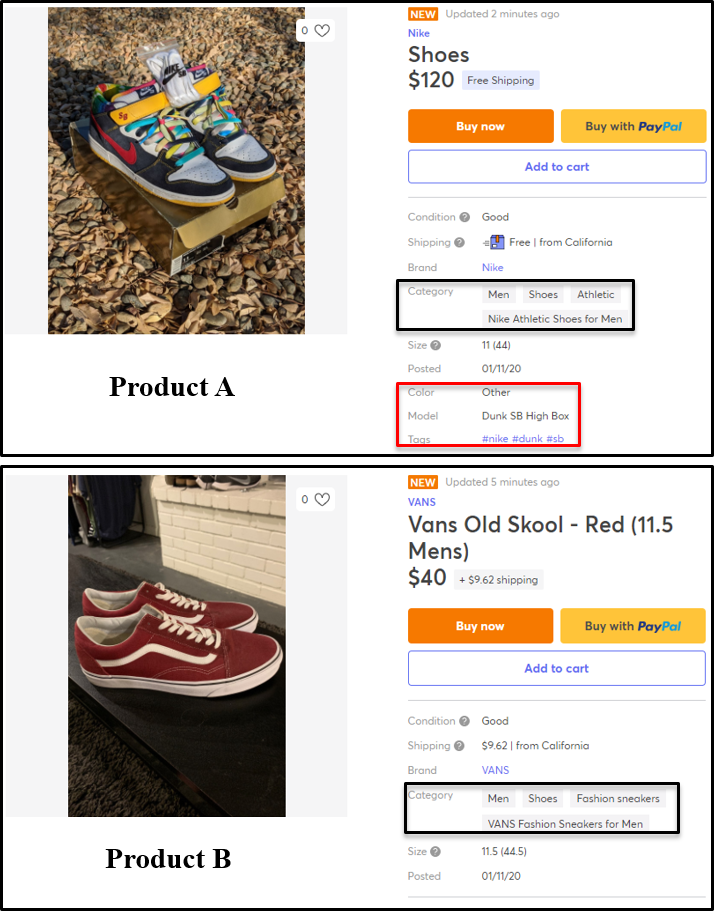}
	\caption{Two items with different attribute information from mercari\textsuperscript{\ref{mecari}}.}
	\vspace{-25pt}
	\label{fig:mecari}
\end{figure}

\begin{figure}[ht]
	\centering
	\includegraphics[width=0.9\linewidth]{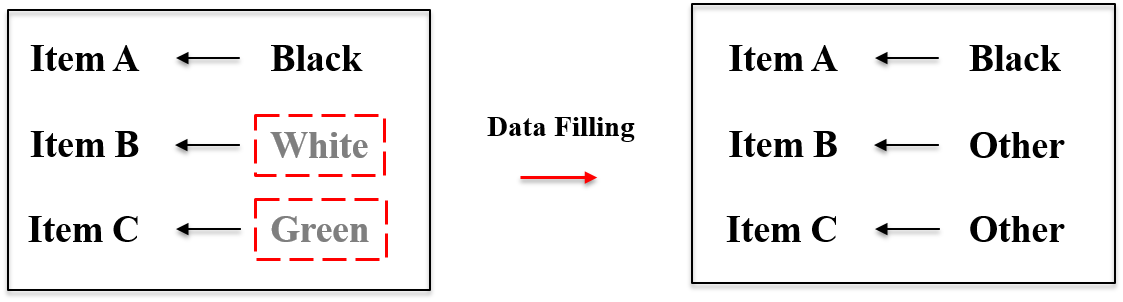}
	\caption{A toy example of substituting the missing attributes with a default value. Red rectangle indicates that the attribute is missing. }
	\vspace{-8pt}
	\label{fig:dfilling}
\end{figure}

Firstly, these methods cannot well tackle the attribute missing problem. It is common that the provided attributes of items are incomplete in real systems. Fig.~\ref{fig:mecari} shows the attributes of two products from Mercari\ref{mecari}, an E-commerce platform. As we can see, each product misses some attributes. Specifically, \emph{Color}, \emph{Model} and \emph{Tags} are missing in Product B. The common solutions to deal with this problem in existing models~\cite{Rendle2011Factorization,he2011NFM,cheng2016wide,shi2018accm} are: 1) \textit{substituting the missing attributes with a default value}~\cite{Rendle2011Factorization,he2011NFM,cheng2016wide} like ``Other", as demonstrated in Fig.~\ref{fig:dfilling}. However, the substituted attribute is meaningless, and the same attribute is used for different items with different real attributes. Consequently, the replaced attribute cannot properly describe the item and will mislead the embedding learning. And 2) \textit{simply assuming that the item does not possess the attribute}~\cite{he2011NFM,cheng2016wide,shi2018accm}, as illustrated by the examples in Table~\ref{table:embedding}. Obviously, this assumption introduces misleading information to the model. Overall, the strategies of the existing models on dealing with missing attributes are inappropriate and may inject biased information to the embedding learning, resulting in sub-optimal performance. 

Another important issue is that most previous methods treat all the attributes equally for both items and users. In those methods, such as NFM~\cite{he2011NFM} and Wide\&Deep~\cite{cheng2016wide}, the embeddings of all attributes are concatenated and projected into a unified feature space without differentiating their importance. We deem that some attributes are more important for an item. What is more, different users may attach importance to different attributes of an item. For example, a young lady may pay more attention to the appearance of a product, such as color and style, while a man may care more about the material or price.

\begin{table}[ht]
	\caption{Illustration of existing methods on the embedding with (E$_{w}$) and without (E$_{w/o}$) an attribute label. The label in red color denotes the missing one.}
	\begin{tabular}{cccc}
		\hline
		Movies            & Tags            & E$_{w}$ & E$_{w/o}$ \\ \hline
		Grumpier Old Men  & Comedy; Romance & $v_{c}v_{r}\overline{v}_{d}$          &  $v_{c}v_{r}\overline{v}_{d}$         \\ 
		Waiting to Exhale & {\color{red}Comedy}; Drama   &  $\bm{v_{c}\overline{v}_{r}{v}_{d}}$          & $\bm{\overline{v}_{c}\overline{v}_{r}{v}_{d}}$          \\ 
		Nixon             & Drama           & $\overline{v}_{c}\overline{v}_{r}{v}_{d}$          & $\overline{v}_{c}\overline{v}_{r}{v}_{d}$          \\ \hline
	\end{tabular}
	\vspace{-8pt}
	\label{table:embedding}
\end{table}

Motivated by the above analyses, in this paper, we present an Attribute-aware Attentive Graph Convolution Network (A${^2}$-GCN for short), which seamlessly incorporates the item attribute information into recommendation\footnote{Our model can be easily extended to consider user attribute information, such as gender and age. In this work, we focus on item attributes, since they are easy to access and more widely used in recommender systems.}. To be more specific, in our model, the users, items and attributes are treated as three types of nodes to construct the graph. Users and items are linked based on their interactions, and attributes are linked to their associated items. The message-passing strategy is used to learn the node representation by aggregating information passed from its neighbor nodes. As to the user representation, it is learned by aggregating messages from neighbor items. Similarly, the item representation of an item is learned by aggregating messages from its neighbor users and attributes. Considering a user may have different preferences for different items, we introduce an attention method to filter the messages passed from different neighbor nodes. Furthermore, the attributes of an item could affect a user's preference for this item. To model this effect, when computing the attention of an item to a user, we design an \textit{attribute-aware attention mechanism} to characterize the attribute information of the given item. As our model only leverages the available attribute information (based on the item-attribute connections in the graph) to learn user and item representations, it naturally avoids the first problem mentioned above. Besides, we devise an attention mechanism to estimate the importance of different attributes, which addresses the second problem in previous methods. Extensive experiments on several large-scale and real-world datasets have conducted to demonstrate the effectiveness of our model by comparing it with several strong baselines.

In summary, the main contributions of this work are as follows:
\begin{itemize}
    \item We step into the existing attribute-aware recommendation models and analyze their shortcomings. Inspired by that, we present a new GCN-based model called A$^2$-GCN, which can naturally address the attribute missing problem in real-world datasets.
    \item We highlight the importance of attribute information on user preference and design a novel attribute-aware attention mechanism to capture the effects of item attribute information on it.
    \item We have conducted extensive experiments on four real-world datasets to demonstrate the superiority of our model over several state-of-the-art baselines. 

\end{itemize}

The rest of this paper is structured as follows. In Section~\ref{sec:related_work}, we briefly survey the related literature. In Section~\ref{sec:methods}, we elaborate the proposed model followed by experiments in Section~\ref{sec:experiments}. We finally conclude this paper in Section~\ref{sec:conclusion}

\section{related work}
\label{sec:related_work}
In this section, we briefly review the recent advancement of model-based collaborative filtering model, especially the attribute-aware and GCN-based methods, which are most close to our work. 

Model-based collaborative filtering methods learn user and item representations based on the user-item interactions for recommendation. This paradigm has achieved great success since the matrix factorization~\cite{Koren2008Fatorization_mn} stood out in the Netflix prize contest~\cite{netflix}. A large amount of research efforts have been devoted into the model-based CF since then and great progress has been achieved thus far, especially the emerging deep learning techniques in recommendation~\cite{he2017neural,wu2016cdae,xue2017deep,He_2018,fouss2007tkde}. Deep learning has been used to learn better user and item representations, due to its powerful capability in representation learning. It is also used to model the complex interaction between users and items. For example, NeuMF~\cite{he2017neural} models the nonlinear interactions between users and items using nonlinear neural networks as the interaction function. 
In addition, metric learning based recommendation methods which use Euclidean distance to model the interactions, have also attracted lots of attentions, because of its better capability of capturing fine-grained user preference over the inner product based approaches~\cite{hsieh2017cml,tay2018latent,liu2018MAML}. 
Although these methods have greatly enhanced the performance, they still suffer from the sparsity problem because they merely rely on the interaction data. A common solution is to leverage side information, such as reviews and attributes, to assist the user and item learning, because side information can provide additionally valuable information. In the next, we mainly discussed the attribute-aware and GCN-based recommendation methods.

\subsection{Attribute-aware Recommendation Models}

Attributes are widely available and valuable for item description. Thus, many methods have been developed to leverage attributes in recommendation~\cite{Rendle2011Factorization,cheng2016wide,Guo2016TKDE,Zhang@TKDE}. Factorization Machines (FMs) based models can model the second-order interactions among attributes via the inner product of their factorized vectors~\cite{Rendle2011Factorization,pasricha2018TransFM,he2011NFM,He2017AFM}. Moreover, they can work with any real-valued feature vectors by modelling all interactions between each pair of features via factorized interaction parameters for prediction.
As the inner product violates the triangle inequality and cannot capture finer-grained attribute interactions, TransFM~\cite{pasricha2018TransFM} employs the squared Euclidean distance function to replace the inner product for sequential recommendation.
The performance of Factorization Machine (FM)~\cite{Rendle2011Factorization} is limited by its linearity and the modelling of pairwise feature interactions. To deal with the linearity of FM, NFM~\cite{he2011NFM} is proposed to  handle the complex and underlying nonlinear structure in real-world dataset. In this method, the attribute information, such as user and item features, is used as side information and embedded into different vectors. AFM~\cite{He2017AFM} extends NFM by adding an attention mechanism to discriminate the importance of different feature interactions.

Another line of research adopts the deep neural network (DNN) techniques to learn the attribute embeddings and concatenate them for recommendation~\cite{cheng2016wide,shanying2016Deepcrossing,shi2018accm,chen2018attention}. 
Wide\&Deep~\cite{cheng2016wide} combines the deep neural network and the linear model for recommendation, where the deep part is a multi-layer perceptron (MLP) on the concatenation of feature embedding vectors to learn feature interactions. DeepCross~\cite{shanying2016Deepcrossing} shares a similar framework with Wide\&Deep by replacing the MLP with the state-of-the-art residual network. ACCM\cite{shi2018accm} adaptively adjusts the source information of each user-item pair with an attention mechanism. 

All the above methods suffer from the two limitations discussed in Section 1. In this paper, we developed a GCN-based attribute-aware recommendation model, which can effectively tackle those limitations.

\subsection{GCN-based Recommendation Models}
In recent years, Graph Convolutional Networks (GCNs) have achieved great success due to the powerful capability on representation learning from non-Euclidean structure~\cite{wang2019ngcf,ying2018pinsage,zheng2018scf,fan2019social,berg2019gcmc,wei2019mm,wang2019kdd}. The main idea of GCNs is how to iteratively aggregate feature information from local graph neighborhoods via neural networks. 

Recently, the GCNs have attracted increasing attention in recommendation. For example, PinSage~\cite{ying2018pinsage} combines random walks with multiple graph convolutional layers on the item-item graph for Pinterest image recommendation. Moreover, the CF effect is captured on the level of item relations, rather than the collective user behaviors. SpectralCF~\cite{zheng2018scf} reveals the proximity information of a graph and the high-order connectivity information via convolution operation in the spectral domain. A drawback of this method is that the eigen-decomposition of graph adjacency matrix causes a high computational complexity. NGCF\cite{wang2019ngcf} exploits high-order proximity by propagating embeddings on the user-item interaction graph. It can simultaneously update the representations for all users and items in an efficient way by implementing the matrix-form rule. GC-MC~\cite{berg2019gcmc} applies the GCN techniques on user-item graph with edges labeled with the observed ratings. It employs one convolutional layer to exploit the direct connections between users and items. 

There are also GCN-based models specially designed for specific recommendation scenarios, such as social recommendation~\cite{fan2019social}, micro-video recommendation~\cite{wei2019mm}, and knowledge-aware recommendation~\cite{wang2019kdd}. In particular, GraphRec~\cite{fan2019social} introduces a method to consider heterogeneous strengths of social relations for social recommendation. GCN-PHR~\cite{wei2019mm} leverages GCN techniques to model the complicated interactions among $<$users, hashtags, micro-videos$>$ and learn their representations. KGAT~\cite{wang2019kdd} models the high-order connectivities in Knowledge Graph based on an end-to-end learning framework.

In this paper, we presented an attribute-aware attentive GCN recommendation model, which leverages attribute information of items to learn better user and item representation for improving recommendation performance. Besides, our model can be used for different scenarios with attribute information, such as E-commerce product, social images, and videos.

\begin{figure}[!]
	\centering
	\includegraphics[width=1.0\linewidth]{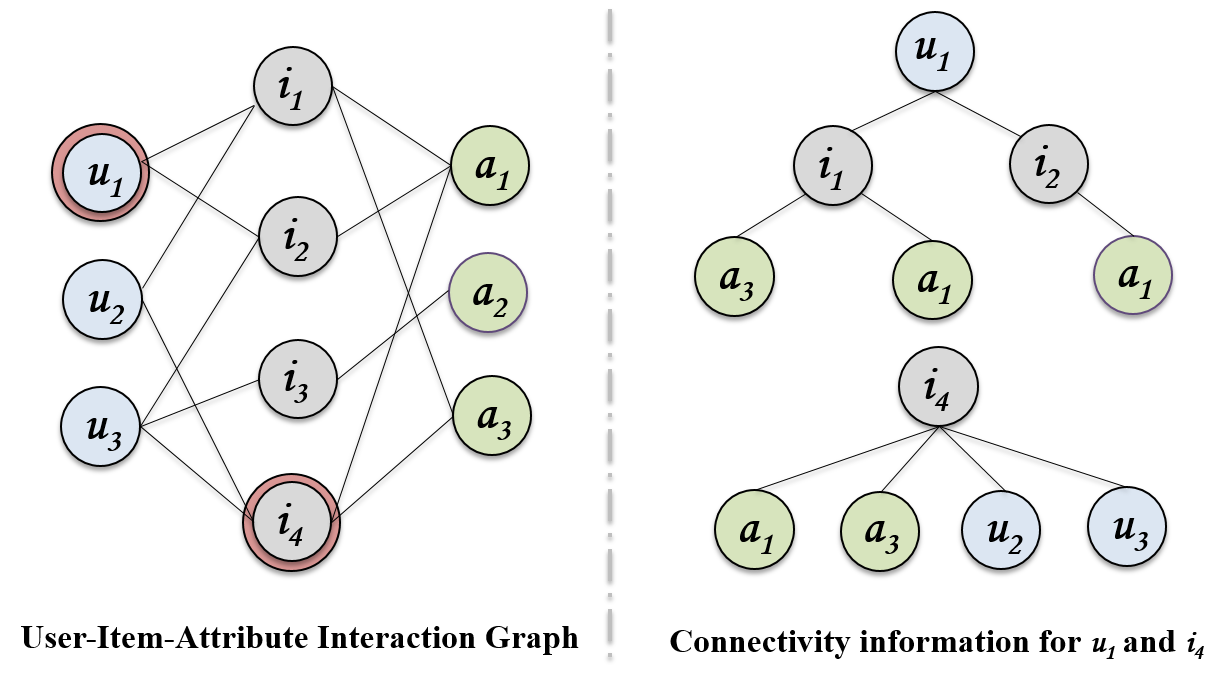}
	\caption{A toy example of an interaction graph based on $<$users, items, attributes$>$ and connectivity information for $u_{1}$ and $i_{4}$. Blue, grey and green circles denote users, items and attributes, respectively. The node $u_1$ and node $i_4$ are respectively the target user and item in this example.}
	\vspace{-8pt}
	\label{fig:tripartite_graph}
\end{figure}
\section{Our Model}
\label{sec:methods}
\subsection{Preliminaries}

Before describing our model, we would like to introduce the problem setting first. Given a dataset with an interaction matrix $\bm{R}^{N_u \times N_v}$ of a set of users $\mathcal{U}$ and a set of items $\mathcal{V}$, where $N_u$ and $N_v$ are the numbers of users and items, respectively. In the dataset, each item is associated with a set of attribute labels $a \in \mathcal{A}$ which describe different attributes of the item. In addition, the number of attribute labels is defined as $N_a$. The matrix $\bm{R}$ records the interaction history between users and items.  A nonzero entry $r_{uv}\in \bm{R}$ indicates that user $u \in \mathcal{U}$  has interacted with  item ${v} \in \mathcal{V}$ before; otherwise, the entry is zero. Notice that the interactions can be implicit (e.g., click) or explicit (e.g., rating). Based on this dataset, the goal is to recommend a user $u\in \mathcal{U}$ with suitable items, which the user did not consume before and will be appealing to.

Let $\mathcal{G}= (\mathcal{W}, \mathcal{E})$ be an undirected graph, where $\mathcal{W}$ denotes the set of notes and $\mathcal{E}$ is the set of edges. Specifically, $\mathcal{W}$ consists of three types of nodes: users $u_{i} \in \mathcal{U}$ with $i \in \{1,\cdots,N_{u} \}$, items $v_{j} \in \mathcal{V}$ with $j \in \{1,\cdots,N_{v} \}$, attributes $a_{k} \in \mathcal{A}$ with $k \in \{1,\cdots,N_{a} \}$, and $\mathcal{U}\cup\mathcal{V}\cup\mathcal{A}=\mathcal{W}$. For the ease of presentation, $i$, $j$ and $k$ will be assigned to index user, item and attribute, respectively. In the graph, user nodes and item nodes are linked based on their interaction history; item nodes and attribute nodes are linked based on the association of attributes to the corresponding items. Note that there is no edge between user nodes and attribute nodes. Fig.~\ref{fig:tripartite_graph} illustrates a toy example of an interaction graph which is based on $<$users, items, attributes$>$ and connectivity information for user $u_{1}$ and item $i_{4}$. We take user $u_1$ and item $i_4$ as examples for illustration.  User $u_1$ has neighbor nodes $i_1$ and $i_2$; $i_1$ has attributes $a_1$, $a_3$ and $i_2$ links to its attribute node $a_1$. For item $i_4$, it has interactions with user $u_2$, $u_3$, as well as attributes $a_1$ and $a_3$.

\subsection{Model Overview}
Typically, model-based CF methods learn the representations of users and items by collectively exploiting the user-item interactions. Here, the attribute provides valuable information about item features, which can be leveraged to capture user preferences. The attribute information of an item can affect a user's preference for the item, because different users may prefer different aspects of an item~\cite{cheng2018aspect} and attributes deliver item information from different aspects. For example, some users like a \emph{movie} because of its \emph{director} while others favor it because of the \emph{leading actor}. Therefore, different users focus on different features of the movie because of its attribute information, or in other words, the attributes of \emph{director} and \emph{actor} pass different information of the movie to different users. Therefore, to accurately model user and item representations, a desired model should capture the complex interactions among users, items, and attributes.

In this paper, we refer to the graph convolutional network (GCN) techniques, because of its powerful representation learning capability, to learn user and item representation in the graph. The core idea behind the node representation learning in GCN is how to recursively aggregate feature information from local graph neighborhoods via neural networks. In our cases, an item node collects information from the connected user nodes and attribute nodes to learn its representation. In particular, given that different users and attributes provide information from different aspects and contribute to an item with different importance, an attention mechanism is designed to distill useful information passed from user nodes and attribute nodes to the item. Similarly, the representation of a user node is learned by distilling information from the connected item nodes which represent the ones this user has consumed before. As discussed above, the attributes of an item can have important effects on a user's preference for this item. To model the effects, we propose a novel attribute-aware attention mechanism which leverages the information of the attribute nodes connected to the items to help learn the attention vector of this item with respect to the target user. To this end, our model can well exploit the interactions among users, items, and attributes for user and item representation learning.

\subsection{Representation Learning}
We adopt the messaging-passing strategy~\cite{berg2019gcmc}  to learn user and item representation. Let $\bm{e_u} \in \mathds{R}^d$, $\bm{e_v} \in \mathds{R}^d$, and $\bm{e_a} \in \mathds{R}^d$ be the embeddings of user $u$, item $v$, and attribute $a$, respectively. $d$ is the embedding size. In the next, we detail the algorithm for user and item representation learning\footnote{ Given that the primary goal of this work is to study the effectiveness of our model on exploiting attribute information for user and representation learning,  our model does not learn the attribute embeddings in this work. However, the embedding of attribute nodes can also be updated in a similar way as user or item nodes do.}.

\subsubsection{Item Representation Learning}
\label{section:item_representation}
For an item $v$, the representation is learned by aggregating all the information passed from the connected users and attributes. The learning process of item representation is described in the following.

\textbf{Message passing.}  
In our model, the representation of an item $v$ is modeled by accumulating the incoming messages from all the user neighbors $u \in \mathcal{N}_u^v$ and attribute neighbors $a \in \mathcal{N}_a^v$, where $\mathcal{N}_u^v$ and $\mathcal{N}_a^v$ are the user neighbor set and attribute neighbor set of item $v$, respectively. Based on the idea of message passing, the information passed from a user neighbor $u \in \mathcal{N}_u^v$ is defined as:

\begin{equation}
\label{equation:message_passing}
\bm{m_{v \leftarrow u}}=\gamma_{u}^v(\bm{W_{1} e_u}+\bm{W_{2}}\left(\bm{e_u} \odot \bm{e_v}\right)),
\end{equation}
where $\bm{m_{v \leftarrow u}}$ is the embedding vector of the message passed from user $u$ to item $v$. $W_{1}, W_{2} \in R^{d^{'} \times d}$ are trainable weight matrices which can distill useful message from $\bm{u_{i}} $. $d^{'}$ is the transformation size. In the message-passing process, we consider the interaction between two nodes via $\bm{e_u} \odot \bm{e_v}$, where $\odot$ denotes the element-wise product. It makes the message dependent on the affinity between $u_{i}$ and $v_{j}$, e.g., passing more messages from the similar items. This strategy can increase the model representation ability and thus boost the performance for recommendation as demonstrated in \cite{wang2019ngcf}. $\gamma_{u}^v$ is a parameter to control how much information to be passed from this user to the item, which  is computed by an attention mechanism that will be introduced later. 

In the same way, the message passed from attribute node $a$ to item node $v$ is obtained by 
\begin{equation}
\bm{m_{v \leftarrow a}}=\gamma_{a}^v(\bm{W_{1} e_a}+\bm{W_{2}}\left(\bm{e_a} \odot \bm{e_v}\right)).
\end{equation}
The notations in this equation are defined in the same manner.

Besides, the original information of the node is also important. To retain the information of $\bm{v}$, we add a self-connection to node $v$ to recycle the information, which is formulated as follows:
\begin{equation}
\bm{m_{v \leftarrow v}}=\gamma_{v}^v\bm{W_{1} e_{v}}.
\end{equation}

\textbf{Message aggregation.}
Aggregating all messages passed from user $u$ and attribute $a$, item $v$’s embedding $\bm{e_{v}}$ is updated by
\begin{equation}
\bm{e_{v}} =\text {LeakyReLU}\left(\bm{m_{v \leftarrow{v}}}+\sum_{u \in {\mathcal{N}^v_{u}}} \bm{m_{v\leftarrow{u}}}+\sum_{a \in {\mathcal{N}^v_{a}}} \bm{m_{v\leftarrow{a}}}\right).
\end{equation}
We use the activation function LeakyReLU~\cite{Andrew2013leaky} to encode both positive and
small negative signals into messages.

\textbf{Attention mechanism.}
We assume that different neighbor nodes (user nodes and attribute nodes) have different influence on item $v$. Inspired by this idea, we design an attention mechanism to estimate the influence of neighbor nodes on items. The influence from user $u$ to item $v$ is formulated as follows:

\begin{equation}
\label{equation:weight_vu}
s^v_u= g\left({\bm{e_{v}}},  \bm{W^v_u {e_{u}}}\right),
\end{equation}
where $\bm{W^v_u}$ is a weight matrix, and $g\left(\cdot\right)$ is a similarity function to measure the similarity of vectors, which represents the relation of passing message from user node to item node. In particular, the cosine similarity function is applied in our work. 

Similarly, the influence of attribute nodes $a$ on item nodes $v$ can be defined as $s_{v a}$:
\begin{equation}
\label{equation:weight_va}
s^v_a= g\left({\bm{e_{v}}},  \bm{W^v_a{e_{a}}}\right).
\end{equation}
The notations in this equation are defined in the same manner as in Eq.~\ref{equation:weight_vu}. Finally, we normalize the weight $s_{v a}$ to obtain the contribution of each user to item representation:
\begin{equation}
\label{equa:item_softmax}
\gamma_{u}^v=\frac{\exp \left(s^v_{u}\right)}{\sum_{v' \in \mathcal{N}^v_{u}} \exp \left(s^v_{u'}\right) + \sum_{a' \in \mathcal{N}^v_{a}} \exp \left(s^v_{a'}\right) + s_v^v},
\end{equation}
where $s_v^v$ is the weight for the self-connection within item $v$.
Similarly, the contribution of each attribute to item representation $\gamma_{a}^{v}$ and the weight of retaining information by self-connection $\gamma_{v}^{v}$ can be obtained in the same way.

\subsubsection{User Representation Learning}
For a user $u$, the representation is learned by aggregating all the information passed from the connected items. The learning process of user representation is the same as that of item representation.
 
\textbf{Message passing.} The message passed from an item $v$ to a user $u$ is
\begin{equation}
\bm{m_{u \leftarrow v}}=\gamma_{v}^u \left(\bm{W_{1} e_v}+\bm{W_{2}}\left(\bm{e_u} \odot \bm{e_v}\right)\right),
\end{equation}
where $\gamma_{v}^u$ is to control the amount of information passed from item $v$ to user $u$. It is computed by the designed attribute-aware attention method described in the next.

\textbf{Message aggregation.} The user representation is updated by aggregating all messages passed from the connected items:
\begin{equation}
\bm{e_u}=\text { LeakyReLU }\left(\bm{m_{u \leftarrow u}}+\sum_{v \in \mathcal{N}^u_v } \bm{m_{u \leftarrow v}}\right),
\end{equation}
where $\bm{m_{u \leftarrow u}}$ is the retained message, and $\mathcal{N}^u_v$ is the set of user $u$'s neighbors.

\textbf{Attribute-aware attention mechanism.} 
Notice that $\gamma_v^u$ can be computed using the same attention mechanism as in the item embedding learning. That is 
\begin{equation}
\label{equ:attention}
s^u_v=g(\bm{e_u}, \bm{{W^u_v}e_v}),
\end{equation}
where $\bm{W^u_v}$ is the weight matrix and $g(\cdot)$ is a similarity function to measure the similarity of vectors. And then $\gamma_{v}^u$ is obtained by the softmax normalization on $s^u_v$:
\begin{equation}
\gamma_{v}^u=\frac{\exp \left( s^u_v\right)}{\sum_{v' \in \mathcal{N}^u_v} \exp \left(s^u_{v'}\right) + s^u_u},
\end{equation}
where $s^u_u$ is the weight of the self-connection (i.e., $\bm{m_{u \leftarrow u}}$).
As aforementioned, the attributes of an item can affect a user's preference for the item, however, this method cannot capture this effect. From the perspective of message passing, a user's representation is learned based on the messages passed from all the interacted items (i.e., item embedding). As the attributes have important influence on user preference, the attributes of an item should be incorporated to the message distilling process from the item to the target user. Based on this consideration, we design an attribute-aware attention mechanism, which incorporates the attribute embeddings  to compute the $\gamma_{v}^u$. Formally, the weight between $v$ and $u$ is estimated by
\begin{equation}
\label{equ:a_attention}
s^u_v= g\left({\bm{e_{u}}},  \bm{W_v^u}\left({\bm{e_{v}}} + \bm{W_a^v\bm{e_a^v}}\right)\right),
\end{equation}
where $\bm{W_v^u}$ and $\bm{W^v_a}$ are weight matrices, and $g\left(\cdot\right)$ is a cosine  similarity function to measure the similarity of vectors. $\bm{e_a^v}$ is a combination vector obtained based on the embeddings of all the attributes associated with the item $v$. Different combination methods can be used to account for the effects of all the related attribute nodes. Here we use a mean-pooling method for its simplicity, namely,
\begin{equation}
 \bm{e_a^v} = \frac{\sum_{a\in \mathcal{N}_a^v}{\bm{e_{a}}}}{|\mathcal{N}_a^v|}.
\end{equation}
To this end, the influence of all the attribute nodes associated with the item is considered in our model in learning the user representation.  

\subsection{Discussion}
As we can see, the attribute information has been integrated in our A$^2$-GCN model for user and item embedding learning based on their connections to items (and second-order user neighbors). It is worth mentioning that our model can naturally avoid the limitations in existing attribute-aware methods, which are either \textit{using a default value  to represent the missing attributes} (e.g., ``Other") or \textit{simply assuming that the item does not possess the properties of the missing attributes } (e.g., ``Comedy" for the movie ``Waiting to Exhale") as discussed in Section 1. In our model, we do not make any assumption on the missing attributes and only take the available attributes into considerations, because only the linked attribute nodes are directly leveraged for the embedding learning. In this way, it can avoid introducing misleading information to the model by inappropriate assumptions, which may hurt the performance. Besides, our model enables the items with missing attributes to benefit from the attribute information propagated from other items via the graph connections. Our experiments validate the superiority of our model over other attribute-aware recommendation models on dealing with the attribute missing problem (see Section~\ref{sec:missingatt}).

\subsection{Model Learning}
\textbf{Prediction.} After learning the user and item representations,   given a user $u$ and a target item $v$, the user preference for this item is predicted by:
\begin{equation}
    \hat{r}_{uv} = \bm{e_{u}}^T\bm{e_{v}}.
\end{equation}

\textbf{Objective function.}
we target at the top-$n$ recommendation, which aiming to recommend a set of $n$ top-ranked items which match the target user's preferences. Similar to other rank-oriented recommendation work~\cite{zhang2017joint,wang2019ngcf}, we adopt the pairwise-based learning method for optimization. Given two user-item pairs: a positive pair $(u, v^{+})$ and a negative pair $(u, v^{-})$, the positive pair indicates that user $u$ has consumed item $v^{+}$ before; and the negative pair means that no interaction exists between $u$ and $v^{-}$. The assumption of pairwise learning is that the user $u$ should prefer $v^{+}$ to $v^{-}$, namely, $\hat{r}_{uv^{+}} > \hat{r}_{uv^{-}}$. 

The objective function is formulated as:
\begin{equation}
    \mathop{\arg\min} \sum_{(\mathbf{u}, \mathbf{v}^+,\mathbf{v}^-)\in{\mathcal{O}}} -\ln\phi(\hat{r}_{uv^+} - \hat{r}_{uv^-}) + \lambda\left\|\Theta\right\|^2_2,
\end{equation}
where $\mathcal{O}=\{(u, v^+, v^-)|(u,v^+)\in\mathcal{R^+}, (u,v^-) \in\mathcal{R^-}\}$ denotes the training set;  $\mathcal{R^+}$ indicates the observed interactions between user $u$ and $v^+$ in the training dataset, and $\mathcal{R^-}$ is the sampled unobserved interaction set. $\lambda$ and $\Theta$ represent the regularization weight and the parameters of the model, respectively. $\phi$ is the sigmoid function. And the $L_2$ regularization is used to prevent overfitting.

\textbf{Model training.}
To prevent the overfitting, we adopt \emph{message dropout} and \emph{node dropout} in A$^2$-GCN. Message dropout randomly blocks the messages passed from one node to another node in the training. It makes the model more independent of edges. Similarly, node dropout randomly removes some nodes, and thus all outgoing messages from those nodes are blocked. This technique enables the embeddings more robust against the presence or absence of particular nodes. They have been successfully applied in previous models~\cite{berg2019gcmc,wang2019ngcf}. The drop ratios of massage dropout and node dropout are empirically tuned in practice.

The mini-batch Adam~\cite{kingma2014adam} is adopted to optimize the prediction model and update the model parameters. Specifically, for a batch of randomly sampled triplets $(u, v^+, v^-)\in \mathcal{O}$, we first learn the representations of those users and items based on the propagation rules, and then update the model parameters by using the gradients of the loss function.

\textbf{Matrix-form propagation rule.}
In order to update the representation for all users and items in an efficient way, we implement the matrix-form of layer-wise propagation rule~\cite{qiu2018deepinf}, which can be described as follows:
  
\begin{equation}
\mathbf{E} = \sigma \left ( \left ( \mathbf{\mathcal{L}} + \mathbf{I} \right )\mathbf{EW_{1}} + \mathbf{\mathcal{L}E}\odot \mathbf{EW_{2}} \right ), 
\end{equation}
where $\mathbf{E} \in \mathcal{R}^{\left( N_u + N_v + N_a \right)\times d}$ is the representations for users, items and attributes; and $\mathbf{I}$ denotes an identity matrix. $\mathbf{\mathcal{L}}$ represents the Laplacian matrix for the $<$user, item, attribute$>$ based graph, which is formulated as:

\begin{equation}
\mathcal{L}=\begin{bmatrix}
\mathbf{0} &\mathbf{Y_{uv}}  &\mathbf{0} \\ 
\mathbf{Y_{vu}} & \mathbf{0} & \mathbf{Y_{va}}\\
\mathbf{0} &\mathbf{0} & \mathbf{0} 
\end{bmatrix},
\end{equation}
where $\mathbf{Y_{uv}} \in \mathcal{R}^{N_u \times N_v}$ is the attention weight matrix which represents the weights from item nodes to each user node, where $\gamma_{v}^u \in \mathbf{Y_{uv}}$. Analogously, $\mathbf{Y_{vu}} \in \mathcal{R}^{N_v \times N_u}$ and $\mathbf{Y_{va}} \in \mathcal{R}^{N_v \times N_a}$ are the attention weight matrices that represent the weights from user nodes to each item node and the weights from attribute nodes to each item node, respectively.

\begin{table*}[ht]
	\centering
	\caption{ Basic statistics of the experimental datasets. \#total$_{attr}$, \#ave.$_{attr}$, \#max$_{attr}$ and ratio$_{attr}$ represent total number of unique attributes, average number of attributes per item, maximum number of attributes associated with items, and the ratio of items with attributes, respectively.}
	\label{tab:data}
	\begin{tabular}{cccccccccl}
		\toprule
		Dataset&\#user&\#item&\#interactions&sparsity&\#attribute&\#total$_{attr}$&\#ave.$_{attr}$&\#max$_{attr}$&ratio$_{attr}$ \\
		\midrule
		Office Products& 4,905& 2,420 & 53,257 &98.10\% &178  &8,205 &3.39 &6 &100.00\%\\
		Clothing & 39,387& 23,033 & 278,676 &99.97\% &253  &130,997 &5.68 &16 &100.00\%\\
		Toys Games & 19,412& 11,924 & 167,596 &99.93\% &404  &26,844 &2.25 &7 & 97.11\%\\ 
		Kindle Store & 68,223& 61,934 & 982,618 &99.98\% &1,438 &423,263 &6.83 &27 &99.80\%\\
		\bottomrule
	\end{tabular}
	\vspace{-10pt}
\end{table*}
\section{Experiments}
\label{sec:experiments}

To validate the effectiveness of our model, we conducted extensive experiments on four public datasets to answer the following research questions.:

\textbf{RQ1:} Does our A${^2}$-GCN model outperform state-of-art methods on the top-$n$ recommendation task?

\textbf{RQ2:} How does the attribute missing problem affect the performance of our A${^2}$-GCN model?

\textbf{RQ3:} Can our model effectively leverage the attribute information to alleviate the sparsity problem?

\textbf{RQ4:} Can our proposed attribute-aware attention mechanism improve the performance? If can, how much improvement can be achieved?

In the next, we first introduce the experimental setup, and then answer the above research questions in sequence.

\subsection{Experimental Setup}

\subsubsection{\textbf{Datasets}}
The public Amazon review dataset\footnote{http://jmcauley.ucsd.edu/data/amazon.}~\cite{mcauley2013hidden}, which has been widely used for recommendation evaluation in previous studies, is adopted for experiments in this work. Four product categories in this dataset are used in our experiments, as shown in Table~\ref{tab:data}. We  pre-processed the dataset to only keep the items
and users with at least 5 interactions. For each observed user-item interaction, we treated it as a positive instance, and then paired it with one negative item which is randomly sampled from items that the user did not consume before. The basic statistics of the four datasets are shown in Table~\ref{tab:data}. As we can see, the datasets are of different sizes and sparsity levels. For example, the \emph{Office Products} dataset is relatively denser than other datasets. Besides, the diversity of datasets is useful for analyzing the performance of our method and the competitors in different situations.

In this work, we focused on the top-$n$ recommendation task, which aims to recommend a set of top-$n$ ranked items that will be appealing to the target user. For each dataset, we randomly selected 80\% of the interactions from each user to construct the training set, and the remaining 20\% for testing. From the training set, we randomly
selected 10\% of interactions as a validation set to tune hyper-parameters. Each user has at least 5 interactions. Therefore, we had at least 3 interactions per user for training, and at least 1 interactions per user for testing.

\begin{table*}[ht]
\caption{Performance of our A${^2}$-GCN model and the competitors over four datasets. The best and second best results are highlighted in bold.}
\centering
\resizebox{\textwidth}{!}{
\begin{tabular}{c|c||ccc||ccc|c||c}
\hline
Datasets                                        & Metrics & BPR-MF & NeuMF & NGCF & NFM & Wide\&Deep & ACCM & A$^{2}$-GCN &Improv. \\ \hline \hline
\multirow{2}{*}{Office Products}                & HR@20      &0.2218	&0.2347	&0.2503  &0.2446	&0.2512	&\textbf{0.2523}	&\textbf{0.2596*} &2.89\%
          \\                            
          & NDCG@20     &0.0952	&0.0997	&0.1077	&0.1041 &0.1082	&\textbf{0.1092}	&\textbf{0.1166*} &6.77\%
         \\ \hline
\multirow{2}{*}{Toys Games} & HR@20       &0.0947	&0.1213	&0.1366		&0.1468	&\textbf{0.1488}	&0.1438 &\textbf{0.1605*} &7.86\%
         \\
    & NDCG@20    &0.0549	&0.0591	&0.0626		&0.0643	&\textbf{0.0656} &0.0634	&\textbf{0.0740*}        &12.80\% \\ \hline
\multirow{2}{*}{Clothing}   & HR@20     &0.0534	&0.0578	&0.0610	&0.0618	&0.0626	&\textbf{0.0667}	&\textbf{0.0775*}         &16.19\%\\
& NDCG@20    &0.0246	&0.0266	&0.0281	&0.0288	&0.0294	&\textbf{0.0313}	&\textbf{0.0350*} &11.82\%
       \\ \hline
\multirow{2}{*}{Kindle Store}                   & HR@20     &0.2987	&0.3324	&0.3413  &0.3506	&\textbf{0.3588} &0.3468	&\textbf{0.3731*} &3.99\%
        \\
& NDCG@20    &0.1469	&0.1607	&0.1659	&0.1745	&\textbf{0.1801} &0.1681	&\textbf{0.1979*} &9.88\% \\ \hline 
\end{tabular}}
	\begin{tablenotes}
		\footnotesize
		\item The symbol * denotes that the improvement is significant with $p-value < 0.05$ based on a two-tailed paired t-test.
	\end{tablenotes}
	\vspace{-10pt}
		\label{tab:results}
\end{table*}

\subsubsection{\textbf{Evaluation Metrics}}
For each user in the test set, we treated all the items that the user did not interact with as negative items. Two widely used evaluation metrics for top-$n$ recommendation are adopted in our evaluation: Hit Ratio and Normalized Discounted Cumulative Gain:
\begin{itemize}[leftmargin=*]
\item \textbf{Hit Ratio (HR)~\cite{deshpande2004hr}}: It is a recall-based metric, measuring whether the test item is in the top-$n$ positions of the recommendation list (1 for yes and 0 otherwise). 
\item \textbf{Normalized Discounted Cumulative Gain (NDCG)~\cite{he2015trirank}}: This metric emphasizes the quality of ranking, which assigns higher score to the top-ranked items by taking the position of correctly recommended into considerations.
\end{itemize}

For each metric, the performance is computed based on the top 20 results. The reported results are the average values across all the testing users. 

\subsubsection{\textbf{Baselines}}
To demonstrate the effectiveness, we compared our proposed A${^2}$-GCN with a set of strong competitors based on different approaches. Among the competitors, BPR-MF, NeuMF and NGCF only use the user-item interactions without considering any side information; NFM, Wide \& Deep, and ACCM leverage the attribute information. We briefly introduce those baselines in below.

\begin{itemize}[leftmargin=*]
\item \textbf{BPR-MF\cite{Koren2009MF}}: It is a classic matrix factorization based method for top-$n$ recommendation. This method only utilizes the user-item interactions and adopts the Bayesian personalized ranking loss for optimization.
\item \textbf{NeuMF\cite{he2017neural}}: It is a state-of-the-art neural collaborative filtering method. This method uses multiple hidden layers above the element-wise and concatenation of user and item embeddings to capture their non-linear feature interactions. 
\item \textbf{NGCF\cite{wang2019ngcf}}: This is a recently proposed GCN-based recommendation method. In particular, this method explicitly encodes the collaborative signal in the form of high-order connectivities by performing embedding propagation in the user-item bipartite graph. It achieves the state-of-the-art performance among the methods which only use user-item interactions.
\item \textbf{NFM\cite{he2011NFM}}: It is a deep neural factorization machine method, which uses Bi-Interaction Layer to integrate both attribute information and user-item interactions.
\item \textbf{Wide\&Deep\cite{cheng2016wide}}: This method jointly trains a wide linear model with feature transformations and a deep neural network with embeddings. In the deep component, we used the same structure as reported in the paper, which has three layers with size 1,024, 512 and 256, respectively.
\item \textbf{ACCM\cite{shi2018accm}}: This is an attention-based model to unify collaborative filtering based and content-based recommendation for both warm and cold scenarios. This model can automatically choose proper attributes to represent the user and the item with attention mechanism. 
\end{itemize}

To ensure fair comparisons, for all the methods using pair-wise learning, each positive instance is paired with a randomly sampled negative user-item instance in the training procedure. We put great efforts to tune hyperparameters of these methods and report their best performance.

\subsubsection{\textbf{Parameter Tuning}}
We implemented our model with TensorFlow~\footnote{https://www.tensorflow.org.} and carefully tuned the key parameters. Specifically, we tuned the initial learning rate $\ell_0$ among $\{0.1, 0.01, 0.001, 0.0001\}$. The coefficient of ${L}_2$ normalization is searched in $\{10^{-5},10^{-4},\cdots, 10^{1},10^{2}\}$, and message dropout and node dropout are tuned in $[0, 0.8]$ with a step size of $0.1$. We optimized all models with the Adam optimizer, where the batch size is fixed at 1,024. In experiments, the dimension of latent vectors (i.e., $\bm{e_u}$ and $\bm{e_v}$) of all methods is set to 64. We used the Xavier initializer~\cite{Xavier2010xavier} to initialize the model parameters. Besides, model parameters are saved in every 10 eopchs and the models will be early stopped if NDCG does not increase for 50 successive epochs.

\begin{figure*}[ht]
	\centering
	\subfloat[HR on Office]{\includegraphics[width=0.25\linewidth]{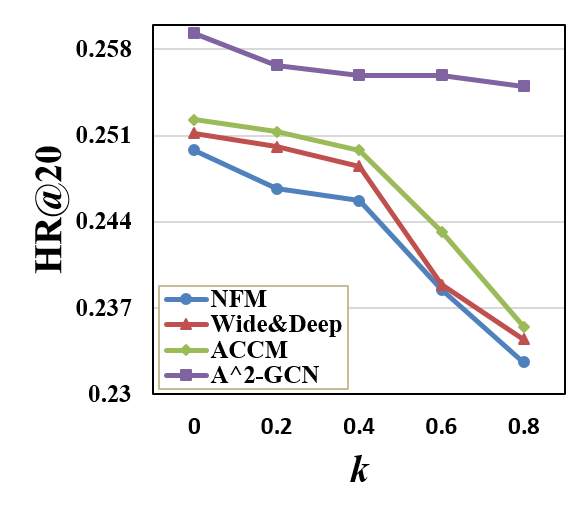}}
	\subfloat[NDCG on Office]{\includegraphics[width=0.25\linewidth]{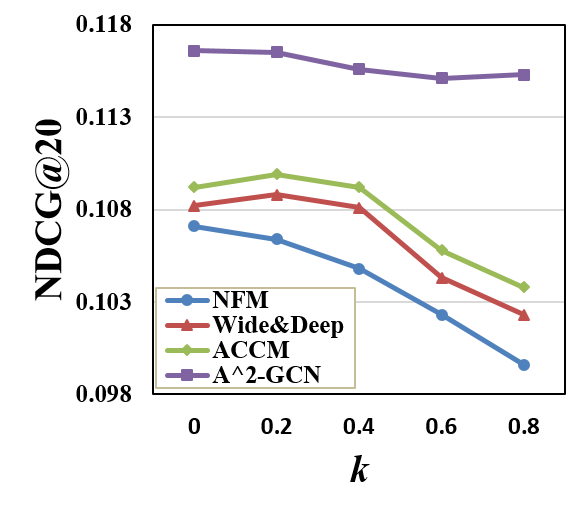}}
	\subfloat[HR on Kindle Store]{\includegraphics[width=0.25\linewidth]{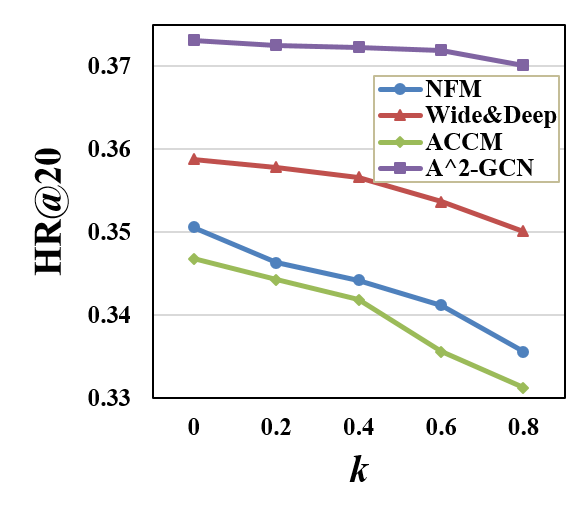}}
	\subfloat[NDCG on Kindle Store]{\includegraphics[width=0.25\linewidth]{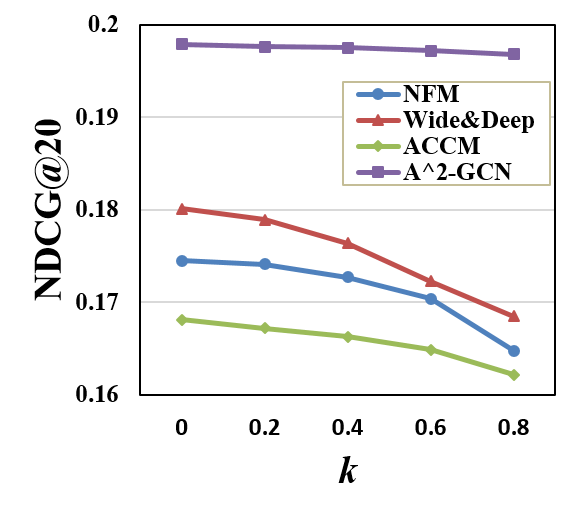}}
	\caption{Impact of attributes missing. Models are trained on training sets with all interactions and attribute labels removed randomly in a ratio of \{0, 0.2, 0.4, 0.6, 0.8\}. $k$ is the  ratio of removed attributes in each dataset.}
	\vspace{-10pt}
	\label{fig:attributes_incomplete}
\end{figure*}
\subsection{Performance comparison (RQ1)}

Table~\ref{tab:results} reports the results of all the considered methods. In the table, both the best and second best performance is highlighted in bold. In addition, we conducted pairwise significant test between our method and the baseline with the best performance. From the results, we have some interesting observations. 

Overall, A$^2$-GCN surpasses all the competitors consistently and significantly across all the cases, demonstrating its effectiveness on leveraging attributes for recommendation. Another interesting observation is that the performance on the \textit{Office Products} and \textit{Kindle Store} is much better than those on the other two datasets. The main reason might be that the item features on the other two datasets are more diverse than that on the \textit{Office Products} and \textit{Kindle Store}. Therefore, it is more difficult to model user preferences on those two datasets, resulting in worse performance. In the next, we analyze the results in detail.

For the methods based on matrix factorization, NeuMF obtains consistently better performance over BPR-MF on all the datasets. The main reason is that the BPR-MF uses the inner products as the interaction function, which cannot capture the complex relations between users and items. In contrast, NeuMF adopts multiple layers of neural networks above the element-wise interaction and concatenation of user and item embeddings to capture the non-linear interactions. This demonstrates the importance of modeling the complicated interactions between users and items. 
NGCF outperforms BPR-MF and NeuMF across all the datasets. This is attributed to the utilization of GCN techniques to learn user and item embeddings on the user-item interaction graph. In addition, NGCF exploits high-order connectivities between users and items through embedding propagation over the graph structure to improve the representation learning. For relatively denser datasets (e.g., \textit{Office Products}),  more neighbor nodes (e.g., interacted items) are available for each node's embedding learning, and thus better performance can be achieved. 

Next, we introduce the performance of the baselines using attribute information (i.e., NFM, Wide\&Deep, ACCM). In general, these methods yield much better performance than those without using attribute information, demonstrating the great value of attributes for recommendation. NFM outperforms NeuMF on all the datasets but underperforms NGCF on the \textit{Office Products}, which might because that \emph{Office Products} is relatively denser. With the rich interactions, NGCF can learn good user and item representations by leveraging high-order proximity information, thus achieving better performance.
Wide\&Deep yields better performance over the above methods on all the datasets.  
ACCM obtains consistently much better performance on \emph{Office Products} and \emph{Clothing}, owing to the adoption of the attention mechanism to control the ratio of information from different sources. This also indicates the importance of differentiating the influence of different information on user/item representation learning. However, it underperforms NFM and Wide$\&$Deep on \emph{Toys Games} and \emph{Kindle Store}. This is because the way of ACCM leveraging attributes makes it only deal with the limited number of attributes. Here, we followed the setting of the ACCM in~\cite{shi2018accm} to consider only the top 100 most frequent attribute labels, which causes great information loss. In contrast, NFM and Wide\&Deep can tackle more attribute labels, and thus yield better performance on the two datasets. 

A${^2}$-GCN outperforms all the baselines consistently over all the datasets. In particular, comparing to the strongest baseline in terms of NDCG@20, A${^2}$-GCN can achieve a relative improvement by 6.77\%, 12.80\%, 11.82\%, 9.88\% on \emph{Office Products}, \emph{Toys Games}, \emph{Clothing} and \emph{Kindle Store}, respectively. The representation learning process of A${^2}$-GCN is the same as NGCF with one embedding propagation layer. The great improvement over NGCF demonstrates the effectiveness of leveraging attribute information in representation learning. Besides, our proposed method yields substantial improvement over the baselines (NFM, Wide\&Deep and ACCM), which also exploit attribute information. This should be credited with the following reasons. Firstly, the GCN technique is used in our method to exploit the attribute information. Due to its powerful capability on representation learning, better performance is expected. Besides, our model can naturally avoid the limitations of those baselines on dealing with the attribute missing problem. This also contributes to the performance improvement (see Section~\ref{sec:missingatt}). Last but not least, our model also benefits from the attention mechanism, which differentiates influence of attributes on user preference in the learning process (See Section~\ref{sec:ablation}).  

Based on the above discussions, we can have following findings. The good performance of NeuMF reveals the importance of modeling non-linear features between users and items. The performance improvement achieved by NGCF on all datasets demonstrates the advantages of GCN-based models and the value of high-order information. Besides, NFM, Wide\&Deep and ACCM indicate the effectiveness of leveraging attribute information to deal with the sparsity problem. 
Our A${^2}$-GCN achieves the best performance because of its effectiveness on exploiting the attribute information by via an attentive GCN method and advantages on tackling the attribute missing problem.

\subsection{Effects of the Attribute Missing Problem (RQ2)}

\label{sec:missingatt}

In this section, we study the influence of the attribute missing problem on attribute-aware recommendation methods. To simulate the attribute missing problem, we randomly removed attribute labels from the datasets by a ratio in \{0, 0.2, 0.4, 0.6, 0.8\}. A ratio of 0.2 indicates that 20\% of the attribute labels of items are randomly removed from the datasets. Due to the
space limitation, we only present the performance on the dataset \emph{Kindle Store} and \emph{Office Products}. The results on the other two datasets show similar trends.

We compare our model with the three baselines using attribute information.  Fig.~\ref{fig:attributes_incomplete} shows the results in terms of HR@20 and NDCG@20 on two datasets. As we can see, A${^2}$-GCN outperforms all the baselines by a large margin, especially when more attribute labels are missing, which indicates the effectiveness of our proposed method on modeling user preferences with attribute information. The performance of three baselines declines rapidly when more attribute labels are missing, especially for the \emph{Office Products}. In particular, the performance of the three baselines deteriorates substantially when $k > 0.4$ on \emph{Office Products}. According to statistics (see Table~\ref{tab:data}), the average numbers of attribute labels per item in \emph{Kindle Store} and \emph{Office Products} are 6.83 and 3.39, respectively. This may explain the dramatical performance drop on \emph{Office Products}. The reasons for performance degradation are two-fold. On the one hand, the removal of attribute labels reduces information source for item and user representation learning. This equally affects all models.  On the other hand, the strategies of those baselines for dealing with missing attributes introduce misleading information into the recommendation model, resulting in inferior performance. 

In contrast, the performance of our model is also negatively affected by the reduction of attribute information, but it is much more stable and the declination is much smaller, especially on the \emph{Kindle Store} because of the relatively rich attribute information. This demonstrates that our model can exploit the attribute information in a much more efficient way via the information propagation through the graph structure. Besides, it also demonstrates the superiority of our model on dealing with the missing attribute problem over the baselines.

\subsection{Effects of Data Sparsity (RQ3)}
\label{sec:sparsity}

\begin{figure}[]
	\centering
	\subfloat[Clothing]{\includegraphics[width=0.85\linewidth]{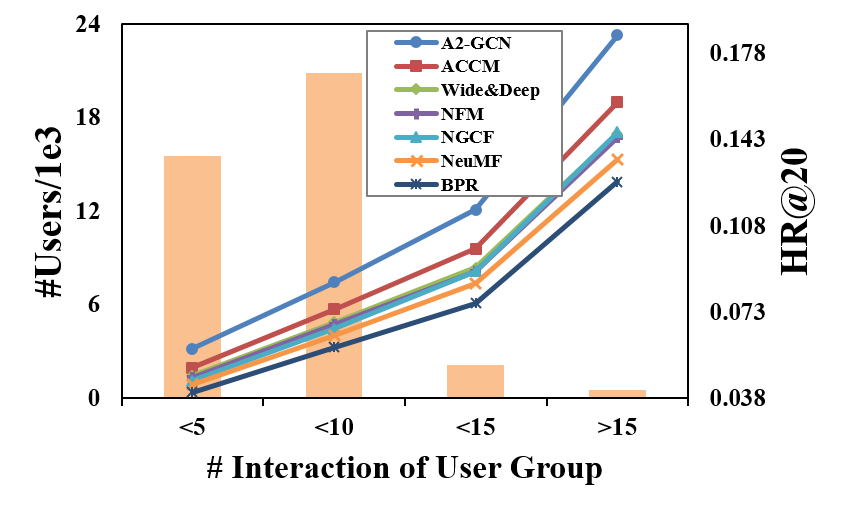}}\vspace{-8pt} \quad
	
	\subfloat[Kindle Store]{\includegraphics[width=0.85\linewidth]{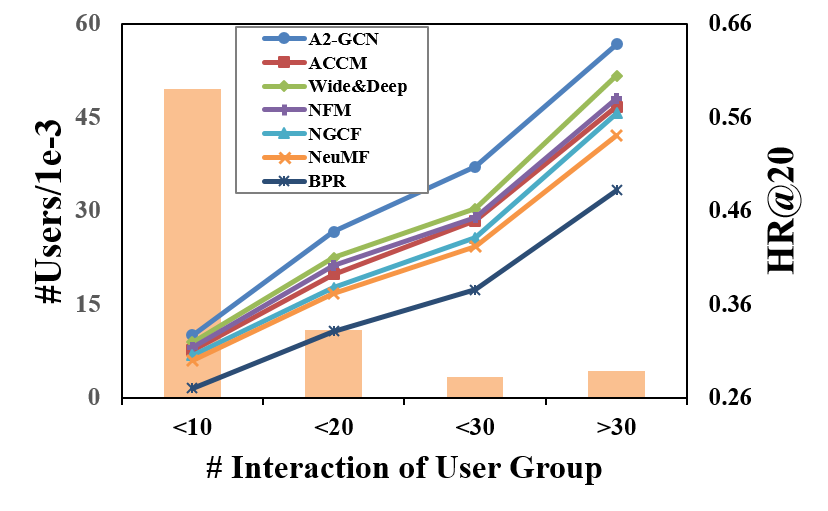}}
	\caption{Performance comparison over the sparsity distribution of user groups on different datasets. Wherein, the background histograms indicate the number of users involved in each group, and the lines demonstrate the performance w.r.t. HR@20.}
	\vspace{-10pt}
	\label{fig:sparsity}
\end{figure}

\begin{table*}[th]
\centering
\caption{Performance of our A${^2}$-GCN model and its variants over four datasets.}
\begin{tabular}{c|c||ccccc}
\toprule
Datasets                         & Metrics &
\begin{tabular}[c]{@{}c@{}}GCN$_{b}$\end{tabular} &
\begin{tabular}[c]{@{}c@{}}A-GCN$_{am}$\end{tabular} &
\begin{tabular}[c]{@{}c@{}}A-GCN$_{att}$\end{tabular} & \begin{tabular}[c]{@{}c@{}}A${^2}$-GCN${_v}$\end{tabular} & A${^2}$-GCN \\ \hline \hline
\multirow{2}{*}{Office Products} & HR@20     &0.2498 &0.2575    &0.2546                                                             &0.2593                                                             &\textbf{0.2596}       \\ 
                                 & NDCG@20   &0.1071 &0.1151   &0.1143                                                             &0.1157                                                             &\textbf{0.1166}       \\ \hline
\multirow{2}{*}{Toys Games}      & HR@20       &0.1347 &0.1505  & 0.1561                                                            & 0.1591                                                            &\textbf{0.1605}       \\
                                 & NDCG@20    &0.0616 &0.0682  &0.0711                                                             &0.0731                                                             &\textbf{0.0740}       \\ \hline
\multirow{2}{*}{Clothing}        & HR@20   &0.0603 &0.0707     &0.0717                                                             &0.0762                                                             &\textbf{0.0775}       \\
                                 & NDCG@20    &0.0274 &0.0323  &0.0332                                                             &0.0343                                                             &\textbf{0.0350}       \\ \hline
\multirow{2}{*}{Kindle Store}    & HR@20      &0.3396 &0.3503  &0.3611                                                             &0.3708                                                             &\textbf{0.3731}       \\
                                 & NDCG@20    &0.1648 &0.1811  &0.1939                                                             &0.1964                                                             &\textbf{0.1979}    \\  
\bottomrule
\end{tabular}
\vspace{-10pt}
\label{tab:variants}
\end{table*}
As important side information, attributes can help alleviate the data sparsity problem. To demonstrate the capability of A${^2}$-GCN for users with limited interactions, we conducted experiments to study the performance of our method and other competitors over user groups with different sparsity levels. In particular, for each dataset, we clustered the users into four groups based on their interaction numbers in the training data. Taking dataset \emph{Clothing} as example, users are divided into four groups based on the number of interactions in the training dataset: less than 5, 10, 15, and more than 15. Fig.~\ref{fig:sparsity} shows the performance in terms of HR@20 on different user groups on dataset \emph{Clothing} and \emph{Kindle Store}. These figures also show the number of users in different groups. We can see that most users have less than 10 interactions in \emph{Clothing} and \emph{Kindle Store}, and many users even have less than 5 interactions in \emph{Clothing}, which further validates the common sparsity problem in real datasets. From the results, we can have the following observations.  
ACCM, NFM and Wide\&Deep consistently obtain better performance than all other baselines. This demonstrates that attribute information is indeed very useful in alleviating sparsity problem. In addition, NGCF surpasses NeuMF and BPR-MF, validating the effectiveness of exploiting collaborative signals from high-order connectivities. A$^2$-GCN yields significantly better performance than all the other baselines over all user groups, which verifies the better capability of our model on exploiting attributes.

By further analyzing the performance improvement on different groups over the two datasets, we can see that the improvement of our model over that of the best baseline becomes larger with the increasing of interactions. Notice that our model benefits from both the attribute information and the GCN technique, which leverages the graph structure to learn user and item representations. With more connections (interactions) in the graph, our model can leverage the graph structure to better exploit the attribute information. It is also interesting to find that our model obtains better performance in the third group over the best baseline than that of other groups in \emph{Kindle Store}. Specifically, relative improvements of A$^2$-GCN over the second best method Wide\&Deep on the four groups are 2.35\%, 6.71\%, 9.76\% and 5.56\%, respectively. The reason might be that, when the interactions are relatively rich, the benefits from attribute and neighbors become limited since interactions provide quit sufficient information for representation learning. 

\subsection{\textbf{Ablation Study (RQ4)}}
\label{sec:ablation}
In this section, we examine the contribution of different components to the performance of our model. Our analyses are based on the performance comparisons to following variants.
\begin{itemize}[leftmargin=*]
\item \textbf{GCN$_b$}: In this model, attribute information and attention mechanism are removed from our method. It is a baseline model which merely applies the GCN technique to the user-item bipartite graph.

\item \textbf{A-GCN$_{am}$}: This variant is designed to investigate the effectiveness of the attention mechanism without attribute information.  

\item \textbf{A-GCN$_{att}$}: This model removes the attention mechanism from our model. This variant is designed to investigate the effects of attributes without attention mechanism. 

\item \textbf{A${^2}$-GCN$_v$}: This model replaces the attribute-aware attention mechanism described in Eq.~\ref{equ:a_attention} by the common mechanism described in Eq.~\ref{equ:attention}, where the attribute information has not been integrated into the attention mechanism. This variant is designed to study the effectiveness of our proposed attribute-aware attention mechanism.  
\end{itemize}

The results of those variants and our method are reported in Table~\ref{tab:variants}. A-GCN${_{am}}$ outperforms GCN${_b}$ over all the datasets, which indicates the importance of differentiating the information distilled from different neighbor nodes in GCN. From the perspective of recommendation, this validates the varying preferences of a user towards different items.
The performance of A-GCN${_{att}}$ indicates the effectiveness of leveraging attribute information for recommendation. In other words, the attributes can provide valuable information in learning better user and item representations. Although A$^2$-GCN$_v$ adopts a simplified attention mechanism, it consistently and substantially outperforms the A-GCN$_{att}$. This also demonstrates the effectiveness of the attention mechanism. More importantly, our A$^2$-GCN further improves the performance based on the proposed attribute-aware attention mechanism. This validates the assumption that items' attribute information affects a user's preference for this item. It also indicates our designed attention model can effectively capture this effect.

\section{Conclusion}
\label{sec:conclusion}
In this work, we present a novel model called Attribute-aware Attentive Graph Convolution Network (A${^2}$-GCN), which can effectively exploit attribute information to learn user and item representations. More importantly, our model can naturally avoid the limitations of previous attribute-aware recommendation methods on dealing with the attribute missing problem. The GCN technique is adopted in A$^2$-GCN to model the complicated interactions among $<$users, items, attributes$>$. In particular, a novel attribute-aware attention mechanism is proposed to capture the effects of item attribute information on user preference. The experimental results on four real-world datasets show that our model can achieve substantially higher recommendation accuracy over several state-of-the-art methods. Additional experiments also validate the effectiveness of A$^2$-GCN on tackling the attribute missing problem and alleviating the sparsity problem.

\ifCLASSOPTIONcompsoc
\else
\fi


\ifCLASSOPTIONcaptionsoff
\fi

\bibliographystyle{IEEEtran}
\bibliography{main}

\end{document}